\documentclass{llncs}
\usepackage{hyperref}
\usepackage{ifpdf}
\ifpdf
 \usepackage[pdftex]{graphicx}
\else
 \usepackage[dvips]{graphicx}
\fi
\usepackage{url}

\pdfoutput=1

\newcommand{\itog}{{\sc i2g}}
\newcommand{\itogatp}{{\sc i2gatp}}
\newcommand{\BibTeX}{{\sc Bib}\TeX}
\newcommand{\BibTeXML}{{\sc Bib}\TeX{\sc ml}}
\newcommand{\XML}{{\sc xml}}
\newcommand{\TGTP}{\textit{TGTP\/}}
\newcommand{\GeoThms}{\textit{GeoThms\/}}

\title{An XML-Format for Conjectures in Geometry\\
(Work-in-Progress)}

\author{Pedro Quaresma
} 
\institute{
  CISUC/Department of Mathematics, University of Coimbra \\
  3001-454 Coimbra, Portugal, \email{pedro@mat.uc.pt} \\
}
\date{}

\begin{document}

\maketitle

\begin{abstract}\textquotedblleft 
  With a large number of software tools dedicated to the visualisation
  and/or demonstration of properties of geometric constructions and also
  with the emerging of repositories of geometric constructions, there is
  a strong need of linking them, and making them and their corpora,
  widely usable.  A common setting for interoperable interactive
  geometry was already proposed, the {\itog\ format}, but, in this
  format, the conjectures and proofs counterparts are missing. A common
  format capable of linking all the tools in the field of geometry is
  missing.  In this paper an extension of the \itog\ format is proposed,
  this extension is capable of describing not only the geometric
  constructions but also the geometric conjectures. The integration of
  this format into the Web-based {\em GeoThms}, {\em TGTP} and {\em Web
    Geometry Laboratory} systems is also discussed.
\end{abstract}

\section{Introduction}
\label{sec:introduction}

In many dynamic \textquotedblleft \textquotedblright geometry software tools (DGSs), a geometric construction is
specified using, explicitly, a formal language. In others, the
construction is made interactively, by clicking specific buttons
and/or icons, but behind this approach there is also a formal
geometric language, although usually hidden from the user. All these
languages share many primitive commands (related to geometric
constructions), but there are also differences in the set of supported
commands, and they follow different syntax rules.

Another important set of tools related to geometric constructions is
given by the geometry automated theorem proving software tools
(GATPs). Given a geometric construction (eventually created with a given
DGS) and a conjecture related to that construction, the GATPs are
capable of proving or disproving (although not always) the conjecture.
Some of them aim at producing traditional, human readable, geometric
proofs~\cite{Chou96,Janicic2010,Narboux04}.

With a large number of tools focusing on visualising geometric
constructions, on proving properties of constructed objects (or both)
and repositories of geometric problems (RGPs), there is an emerging
need of linking them and making widely usable: constructions; conjectures
and proofs generated with different tools. This would help in the
progress of the field of geometric constructions, including their role
in education.

The \itog\ format~\cite{Santiago2010} was designed to describe
constructions created with a DGS allowing the exchange of geometric
constructions between different DGSs. This format should be complemented
in such a way that it can provide support for conjectures. The new
format should be a superset of the former format, i.e. a DGS should be
able to read the new format, ignoring all
the extra information regarding conjectures and proofs. A GATP should be
able to read the new format using, if needed, the geometric construction
specification. In the following such an extension, the \itogatp\ format,
is discussed.

Some of the most important motivating arguments for using {\scshape
  xml} in storing descriptions of geometric constructions and
conjectures and as an interchange format are: strictly structured
files, easy to parse, process, and convert into different forms and
formats; a strict content validation of documents with respect to a
given set of restrictions; easier communication and exchange of
material between unrelated tools.

%%%%%%%%%%%%%%%%%%%%%%%%%%%%%%

% \begin{itemize}
% \item Instead of raw, plain text representation, geometric
%  constructions and conjectures will be stored in strictly
%  structured files; these files will be easy to parse, process, and
%  convert into different forms and formats.
% \item Input/output tasks will be supported by generic, external tools
%  and different geometric tools will communicate easily.
% \item Repositories of geometric constructions will be unified and
%  accessible to users of different geometric tools.
% \item Easier communication and exchange of material with the rest of
%  mathematical and computer science community.
% \item There is a wide and growing support for {\scshape xml}.
% \item Different sorts of presentation (plain text, \LaTeX, {\scshape
%  html}) easily enabled.
% \item Strict content validation of documents with respect to a given set
%  of restrictions.
% \end{itemize}

%%%%%%%%%%%%%%%%%%%%%%%%%%%%%%

\paragraph{\bf Paper overview.} In Section~\ref{sec:background} 
some background regarding DGSs, the \itog\ format, GATPs and RGPs is
given. In Section~\ref{sec:overallarchitecture} the overall structure
of the new format is described. In Sections~\ref{sec:implementation}
implementations issues are discussed. Finally in
Section~\ref{sec:conclusions} some final conclusions are drawn and
future work is discussed.

% -----------------------------------------------

\section{Background}
\label{sec:background}

In this section some basic background information about geometric
constructions, the intergeo format (\itog), geometric conjectures and
proofs and repositories of geometric problems is given.

\subsection{Dynamic Geometry Software}
\label{sec:dgs}

Dynamic geometry software tools (DGSs) allow an easy construction of
geometric figures built from free objects, elementary constructions and
constructed objects.  The dynamic nature of such tools allows its users
to manipulate the positions of the free objects in such a way that the
constructed objects are also changed, yet preserving the geometric
properties of the construction. These manipulations are not formal
proofs, as the user is considering only a finite set of concrete
positions. Neither the DGS are able to provide a proof of a given
conjecture nor they are able to ensure the soundness of the
constructions built by its users.

%~\footnote{By soundness of a construction I mean that the
%  construction is valid and the elements in it are (geometrically) true.}

% soundness: a property of both arguments and the statements in them,
%i.e., the argument is valid and all the statement are true.

There are multiple DGSs available~\footnote{\url{www.geogebra.org},
  \url{www.cinderella.de}, \url{www.dynamicgeometry.com/},
  \url{zirkel.sourceforge.net/}, \url{www.cabri.com/},
  \url{www.emis.de/misc/software/gclc/}}: GeoGebra, Cinderella,
GeometerSketchpad, C.a.R., Cabri, GCLC to name some of the most used.

\subsection{Intergeo Format}
\label{sec:i2g}

The Intergeo (\itog) file format is a specification based on the markup
language \XML\ designed to describe constructions created with a DGS. It
is one of the main results of the intergeo project, an eContentplus
European project dedicated to the sharing of interactive geometry
constructions across boundaries. For more information about the
project, visit the site \url{http://i2geo.net} and look into the
documentation available there, as well as
to~\cite{Kortenkamp2006,Kortenkamp2009}.

An intergeo file takes the form of a compress file package. The main
file is {\tt intergeo.xml}, which provides a textual description of
the construction in three parts, the elements part describing a
(static) initial instance of the configuration, the constraints part
where the geometric relationships are expressed and the display part
where the details regarding the rendering of the construction are
placed.  For more details on the file format see~\cite{Santiago2010}.

There are already a significant number of DGSs supporting the \itog\
format (see~\cite{i2gImplementationTable} for details). 

\subsection{Geometry Automated Theorem Proving}
\label{sec:gatp}

The geometry automated theorem provers (GATPs) give its users the
possibility to reason about a given DGS construction, this is no longer
a ``proof by testing'', but an actual formal proof. Another link between
the GATPs and the DGSs is given by the automated deductive testing, by
the GATP, of the soundness of the constructions made by the
DGS~\cite{Janicic06b}.  Most, if not all, DGSs are capable of detecting
and reporting syntactic and semantic errors, but the verification of the
soundness of the construction is beyond their capabilities. If we can
link DGSs and GATPs we will be able to use a given GATP in order to
check the soundness of a construction created with the help of a DGS.

% acrescentar referencias a v\'arios GATPs --- IJCAR? ou um desses

Automated theorem proving in geometry has two major lines of research:
synthetic proof style and algebraic proof style (see~\cite{Matsuda04}
for a survey). Algebraic proof style methods are based on reducing
geometric properties to algebraic properties expressed in terms of
Cartesian coordinates.  These methods are usually very efficient, but
the proofs they produce do not reflect the geometric nature of the
problem and they give only a yes/no conclusion.  Synthetic methods
attempt to automate traditional geometry proof methods producing
human-readable proofs.

If the GATP is capable of producing synthetic proofs, the proof itself
can be an object of study, in other cases only the conclusion
matters~\cite{Chou96,Janicic06a}.

\subsection{Repositories of Geometric Problems}
\label{sec:repositoriesgeometricproblems}

When considering repositories of geometric problems we are directly
interested in a common format. If we want to provide a repository of
geometric problems that can be used by DGSs and GATPs, then the
constructions should be kept in a common format that can be converted to
the DGS and/or GATP internal format whenever needed. The author of this
paper is directly involved in this efforts having three different
project that involve repositories of geometric problems.

The first (chronologically) of the mentioned projects is
GeoThms\footnote{\url{http://hilbert.mat.uc.pt/GeoThms/}}, a Web-based
framework for exploring geometric knowledge integrating DGSs, GATPs,
and a RGP.  The GeoThms is a publicly accessible system with a growing
body of geometric constructions and formally proven geometric
theorems, its users can easily use/browse through existing geometric
contents and build new contents~\cite{Quaresma2007}. Within this
project a common, {\scshape xml}-based, interchange format for
descriptions of geometric constructions, conjectures and proofs was
developed~\cite{Quaresma08}. This format predates the \itog\ format.

A more recent project is the Thousands of Geometric problems for
geometric Theorem Provers
({\TGTP})\footnote{\url{http://hilbert.mat.uc.pt/TGTP}}. This is a
Web-based library of problems in geometry. {\TGTP} aims, in a similar
spirit of {\em TPTP} and other libraries, to provide the automated
reasoning in geometry community with a comprehensive and easily
accessible library of GATP test problems~\cite{Quaresma2011}. The
\itogatp\ format is being developed for this project. For the moment the
\TGTP\ system still uses the {\scshape xml}-based, interchange format
developed for the GeoThms system (the two system share a common
database), but it will change to the new format as soon as it becomes
stable.

In an educational setting, the project Web Geometry Laboratory
(WGL)\footnote{In prototype stage:
  \url{http://hilbert.mat.uc.pt/WebGeometryLab/}} is an
asynchronous/synchronous Web environment that integrates a DGS and a RGP
(and it will integrate a GATP in a next version), aiming to provide an
adaptative and collaborative blended-learning environment for
geometry~\cite{Santos2012}. Here the need for a common interchange
format is less important, nevertheless it will be useful to allow the
system to be more easily configurable, i.e. using a common format will
allow choosing the DGS and/or the GATP more freely.

% ------------------------------------------------------------------------

\subsection{Integration Issues}
\label{sec:integrationissues}

There are already some systems integrating a DGS with one, or more, GATP
and a set of examples (e.g.  GCLC~\cite{Janicic06c,Janicic06a},
GeoProof~\cite{Narboux2007}, JGEX~\cite{Chou2011}), but all this systems
provide closed tools with a tight integration between different internal
functionalities.  If we want to be more generic, loosely linking DGSs,
GATPs and RGPs, we need a way to establish the communication between
tools as unrelated modules, i.e. we need a common format that can be
used as a communication channel between tools.

%%%%%%%%%%%%%%%%%%%%%%%%%%%%%%%%%%%%%%%%

\section{Overall Architecture}
\label{sec:overallarchitecture}

A common format for geometric constructions, conjectures and proofs
should address the communication between DGSs and GATPs, to establish
the soundness, by the GATP, of a construction made with the help of
the DGS or to prove (or disprove) a given conjecture about a
construction made in the DGS:

%%%%%%%%%%%%%%%%%%%%%%%%%%%%%%%%%%%%%%%%

\begin{itemize}
\item The communication between DGSs and GATPs, to establish the
 soundness, by the GATP, of a construction made with the help of the
 DGS or to prove (or disprove) a given conjecture about a
 construction made in the DGS.
\item The rendering of the proof. If the GATP uses an algebraic method
 only the final result will be usable, but if the GATP uses a
 synthetic method, the proof itself can be an object of study.
\end{itemize}

%%%%%%%%%%%%%%%%%%%%%%%%%%%%%%%%%%%%%%%%

Geometric proofs could appear in many different forms, for instance in
axiomatic form (e.g., in Hilbert-style, sequent calculus style, etc.);
representing higher-level proofs, produced by the area method; as
algebraic proofs produced by the algebraic methods like the Gr\"obner
basis method, etc. The representation of the proof and/or its
rendering will always be linked to the method used in its
development. This will be addressed in Section~\ref{sec:formatXML}.

\subsection{Representation of Constructions, Conjectures and Proofs}
\label{sec:constructions}

In order to enable communication between the geometric tools (i.e DGSs
and GATPs) and converting files between different formats a single
target format should exist: a format that could define a common normal
form for the different tools.  The proposal is to extend the \itog\
format in such a way that the new format would complement the
construction description (made by a DGS) with the conjecture
description. This new format will be called the \itogatp\ format.

Converting from a DGS/GATP language to {\scshape xml}, would be
performed by a specific converter, naturally relying on the DGS/GATP's
parsing mechanism. Converting from {\scshape xml} to a DGS/GATP
language, will be implemented via \XML-parsing tools. 

Having converters from, and to, the \itogatp\ format for all DGSs and
GATPs, we (indirectly) have converters from each tool to any other
tool.  Thus, in this way, the base for a common interchange format is
provided.  {\scshape xml} is a natural framework for such interchange
format, because of its strict syntax, verification mechanisms,
suitable usage on the Internet, and a large number of available
supporting tools.

{\scshape xml} descriptions of constructions, conjectures and proofs
can be, by means of {\scshape xslt}, also rendered into other formats
that are convenient for human-readable display in browsers.  It can
also be transformed into different representations, such as natural
language form.

A specific {\scshape xml} scheme document could define syntactical
restrictions for construction descriptions, conjectures and proofs.
This document could then be used, in conjunction with the generic
{\scshape xml} validation mechanism, for verifying whether a given
file in the \itogatp\ format is correct (or not).

\subsection{Structure of \itogatp\ Format}
\label{sec:formatXML}

\begin{figure}[hbtp]
 \centering
   \ifpdf
   \includegraphics[width=1\textwidth]{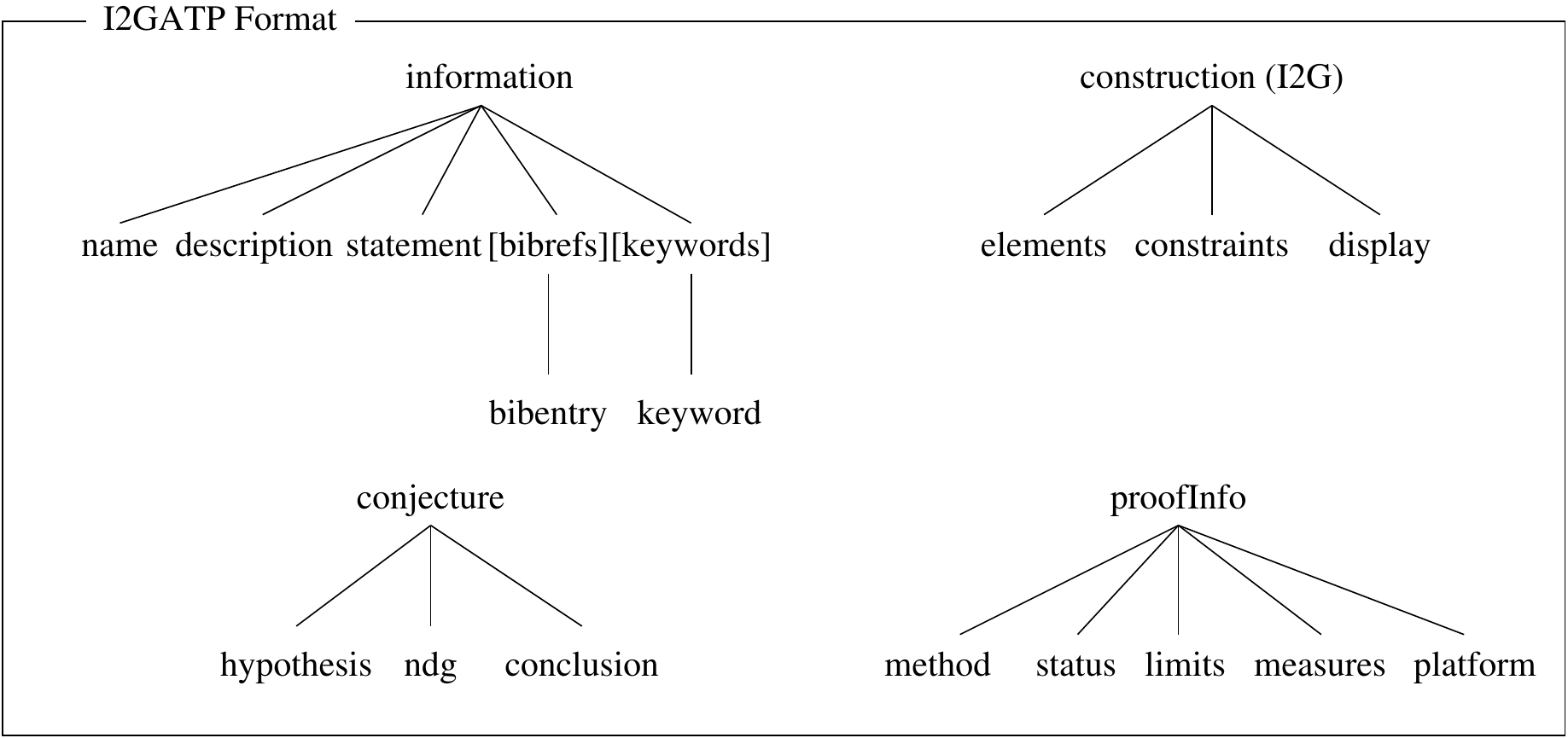}
   \else
   \includegraphics[width=1\textwidth]{formatXML.eps}
   \fi
 \caption{Structure of the {\itogatp} File Format}
 \label{fig:structureI2GATPfileformat}
\end{figure}

Following the ideas of the \itog\ common format all the files related
to the \itogatp\ format will be packed in a single compressed file,
the {\em container}, which is nothing more then a \itog\ container
with three additional directories. The \itogatp\ format will be spread
in four, at least, {\sc XML} files (see Figure~\ref{fig:structureI2GATPfileformat}).
%%%%%%%%%%%%%%%%%

Apart from the {\tt intergeo.xml} file, which is mandatory (see the
\itog\ format specification~\cite{Santiago2010}), the other files are
optional. 

%%%%%%%%%%%%%%%%%%%%%%%%%%%%%%%%%%%%%%%%

%%%%%%%%%%%%%%%%%%%%%%%%%%%%%%%%%%%%%%%%

\paragraph{Information}
\label{sec:infosubtree}
The {\tt information.xml} file contains all the generic (human)
information about the problem. The {\em name} of the problem; a brief,
informal, {\em description} of the problem; an informal (rigorous)
mathematical description ({\em statement}) of the problem; a list of
bibliographic references; a list of keywords.

\paragraph{Construction}
\label{sec:constructionsubtree}
The {\tt intergeo.xml} file contains the construction in the \itog\
format. The \itog\ format has as main tag the {\em construction} tag
with three sub-nodes: {\em elements} for the free objects; {\em
  constraints} for the objects fixed by construction constraints and
{\em display} for the display details.

%In the next section we will speak about the {\em container} which will
%be a superset of the \itog\ container.

\paragraph{Conjecture}
\label{sec:conjectureSubtree}

This is the core of the \itogatp\ format. In here the {\em
  hypothesis}, the {\em ndg} (non-degenerate conditions) and the {\em
  conclusion}, establishing the conjecture to be proved, are
specified. The non-degenerate conditions could be a side-effect of the
proving process, e.g. automatically generated by a GATP based in the
area method, or provided manually.

\paragraph{Proofs}
\label{sec:proofsSubtree}

For a given problem/conjecture we can have many proof attempts:
different approaches, for instance synthetic proof versus algebraic
proof; different methods, Gr\"obner bases method versus Wu's method;
different GATPs, GCLCprover versus CoqAM, and all the possible
combinations of this three different aspects. 

Each proof attempt will be kept in a file {\tt proofInfo.xml} in a
sub-directory of the proofs directory (see Section~\ref{sec:container}
for more details).

Each individual proof node will have: the information regarding the
GATP, its version and method used; the status of the proof, e.g. {\em
  proved}; the computational constraint regarding the proof attempt
made by the GATP, e.g. maximum CPU time and RAM space allowed by the
system; the proof metrics, e.g. number of proof steps (area method)
and the platform used when doing the proof, e.g. CPU, RAM, and other
details about the computational platform.

For the proof status the SZS ontology~\cite{Sutcliffe2008} will be used
as a base. The ``Unsolved'' branch will be used as it is, the ``Solved''
branch has to be adapted to the \itogatp\ settings.

Given the fact that the proofs produced by different GATPs/Methods
are, and should continue to be, quite different we do not try to
create a common formats for the proofs. The outcomes produced by the
different GATPs will be kept as they are produced (see the {\em
  container} in Section~\ref{sec:container}). 

\subsection{The container}
\label{sec:container}

As said above, the \itogatp\ {\em container} is a superset of the
\itog\ container, with three additional directories: ({\tt information};
{\tt conjecture} and {\tt proofs}). This means that it will be
possible to extract the \itog\ container out of this file, it will be
a simple question of unpacking the file, erasing the additional
directories and repacking, if needed, the resulting files.

%%%%%%%%%%%%%%%%%%%%%%%%%%%%%%%%%%%%%%%%%

\begin{table}[hbtp]
 \centering
 \begin{tabular}{l|c}
   information/ & mandatory \\
   information/information.xml & optional \\
   construction/ & mandatory \\
   construction/intergeo.xml & mandatory \\
   construction/preview.pdf & optional \\
   construction/preview.svg & optional \\
   construction/(\ldots)\\
   conjecture/  & mandatory \\
   conjecture/conjecture.xml & optional\\
   proofs/  & mandatory \\
   proofs/proof$<$GATP$><$Version$><$Method$>$/ & optional \\
   proofs/proof$<$GATP$><$Version$><$Method$>$/proofInfo.xml & optional \\
   proofs/proof$<$GATP$><$Version$><$Method$>$/proofOutput.pdf & optional \\
   proofs/proof$<$GATP$><$Version$><$Method$>$/(\ldots)  \\
   metadata/ & optional \\
   metadata/i2g-lom.xml  & optional \\
   resources/  & optional \\
   resources/$<$image\_files$>$ & optional \\
   resources/(\ldots) \\
   private/ & optional \\
   private/$<$domain-name$>$ & optional \\
   private/$<$domain-name$>$/$<$files$>$ & optional  
 \end{tabular}
 \caption{The \itogatp\ container}
 \label{tab:itogatpcontainer}
\end{table}

%%%%%%%%%%%%%%%%%%%%%%%%%%%%%%%%%%%%%%%%%

The structure of the container follows closely the structure of the
\itogatp\ format. The {\tt information}, {\tt construction} and {\tt
  conjecture} directories will contain the files {\tt
  information.xml}, {\tt intergeo.xml} and {\tt conjecture.xml}
respectively. The directory {\tt construction} may also contain the
rendering of the construction in various graphical formats (e.g. PDF,
SVG, PNG, etc.).

The directory {\tt proofs} will contain as many sub-directories as
proofs attempts were made for the problem in question. The naming
convention follows the ideas in the \itog\ format, that is, after the
prefix ``proof'', the name of the GATP, its version and finally the
method used. Given the fact that this is a directory identifier the
strings used in these last fields should be conform to the standard
naming conventions. In each of this sub-directories the file {\tt
  proofInfo.xml} will contain the information regarding the proof
attempt. This directory may also contain files with the rendering of
the proof in different formats (e.g. PDF, HTML, etc.).

%%% NOTES
%a questão dos nomes dos directorios tem de ser vista com cuidado
%%% end NOTES

The remaining directories follow the structure of the \itog\ format and
can be used to place additional contents produced by the GATPs.  

Following the \itog\ conventions, the suggest naming convention to the
container is {\tt problem$<$problem\_name$>$.zip}.

In the next section the symbol lists, i.e. the tags proposed to this
\XML-format, are described.

\subsection{Symbol Lists}
\label{sec:symbol-lists}

As said above, the container will have ``four'' (main) \XML\ files:
{\tt information.xml}; {\tt intergeo.xml}; {\tt conjecture.xml} and as
many {\tt proofInfo.xml} files as proof attempts were made for a given
problem. The {\tt intergeo.xml} is described in the \itog\ common file
format, technical report D3.10~\cite{Santiago2010}. The other three
are specific for the \itogatp\ format and their symbol lists will be
described in the next sections.

The symbol lists will be describe in a coarse fashion. For a more
detailed account see~\cite{Quaresma2012a}.

\subsubsection{Generic Information ({\tt information.xml})}
\label{sec:infosymbollist}

Generic information about the problem. All fields, except the {\em
  name}, may be empty. 

The tags are: {\em name}; {\em description}; {\em statement}; {\em
  bibrefs} and {\em bibentry}; {\em keywords} and {\em keyword}.

The {\em description} will be a brief, informal, description of the
problem in text format and the {\em statement} will be an informal
(rigorous) mathematical description of the problem in {\sc
  MathML}~\cite{mathml2010}.
%%%%%%%%%%%%%%%%%%%%%%%%%%%%%%%%%%%%

The {\em bibrefs} is a list (it may be empty) of bibliographic
references in \BibTeXML\
format\footnote{\url{http://bibtexml.sourceforge.net/}}.

The contents of the {\em description} and {\em bibrefs} tags could be
automatically converted from \LaTeX\ and \BibTeX\ using, for example,
{\em tex4ht\/}\footnote{\url{http://tug.org/applications/tex4ht/}} and
\BibTeXML\ converters respectively.

The {\em keywords} is a list of keywords in text format. For the
moment this field is a free-form text field. For better querying the
repositories, an index or a geometric ontology should be
considered. Maybe an ``open classification'', that is, a
classification index open to users additions and where the most chosen
keywords became, in time, fixed.

\subsubsection{Conjecture Information {\tt conjecture.xml}}
\label{sec:conjecturesymbollist}

The main tags are: {\em conjecture}; {\em hypothesis}, {\em ndg} (for
non-degenerate conditions) and {\em conclusion}. The three
last tags can contain a large number of other tags used to write down
the geometric (logical) statements.

%%%%%%%%%%%%%%%%%%%%%%%%%%%%%%%%%%%%

Without pretending to be exhaustive we have: {\em not\_equal}; {\em
 not\_parallel}; {\em equal}; {\em plus}; {\em mult}; {\em collinear};
{\em perpendicular}; {\em parallel}; {\em midpoint}; {\em same\_length};
{\em harmonic}; {\em segment\_ratio}. The symbols of the intergeo format
regarding the geometric construction can occur here.

%%%%%%%%%%%%%%%%%%%%%%%%%%%%%%%%%%%%

\subsubsection{Proofs Information {\tt proofInfo.xml}}
\label{sec:proofssymbollist}

Contains all the information regarding a proof attempt for given problem.

This is a record of the conditions under which the proof was
attempted, i.e.  the method used ({\em method}), the limits imposed
to the GATP and the computer system used ({\em limits} and {\em
  platform}). Adding to this the proof outcome, i.e.  proved, not
proved, etc. and also, measures of efficiency, e.g. CPU time used,
number of steps, etc. ({\em status} and {\em measures}).

%%%%%%%%%%%%%%%%%%%%%%%%%%%%%%%%%%%%

In the list of symbols we have (among others): {\em status}; {\em
 limits}; {\em time\_limit\_se\-con\-ds}; {\em iterations\_limit}; {\em
 measures}; {\em CPU\_time}; {\em elimination\_steps}; {\em
 number\_terms\_lar\-gest\_poly\-nomial}; {\em computer\_name}; {\em
 clock\_speed}; {\em RAM}; {\em operating\_system}.

%%%%%%%%%%%%%%%%%%%%%%%%%%%%%%%%%%%%

\section{Implementation}
\label{sec:implementation}

\begin{figure}[hbtp]
 \centering
   \ifpdf
   \includegraphics[width=1\textwidth]{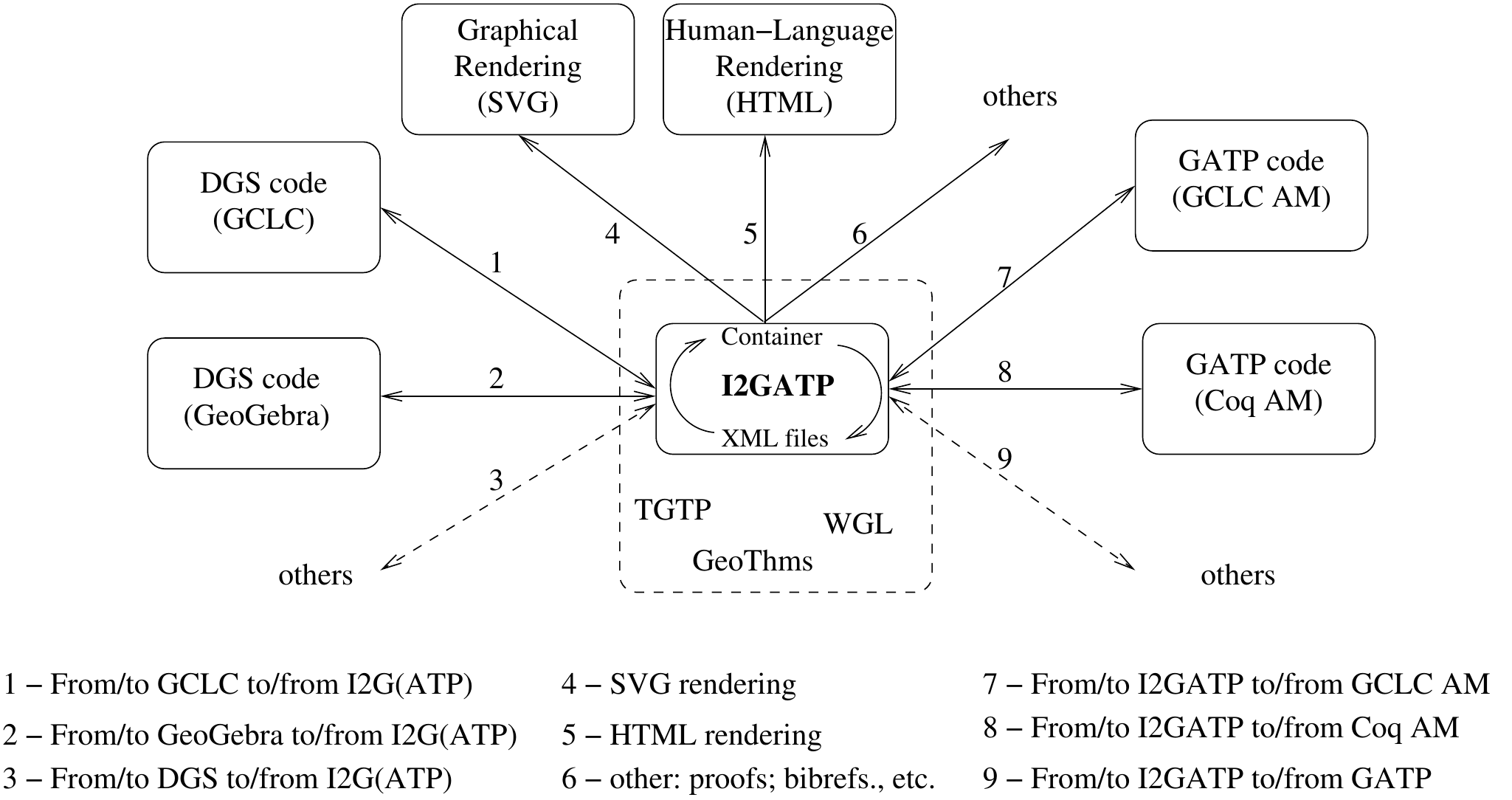}
   \else
   \includegraphics[width=1\textwidth]{convertionsXMLTools.eps}
   \fi
   \caption{Conversions From/To \itogatp\ To/From Geometric Tools}
   \label{fig:convertionXMLTools}
\end{figure}

Having defined a \XML\ format for geometric constructions and
conjectures its usefulness depends on its support from other tools,
i.e. the capability of tools such as DGSs
(see~\cite{i2gImplementationTable} to the list of tools already
supporting the \itog\ format) and GATPs to export to the
\itogatp\ format and, of course, its support to other tools in the shape
of converters from \itogatp\ format to the internal format of tools such
as the DGSs and GATPs (see Figure~\ref{fig:convertionXMLTools}).

%%%%%%%%%%%%%%%%%%%%%%%%%%%%%%%%%%%%%%%

%%%%%%%%%%%%%%%%%%%%%%%%%%%%%%%%%%%%%%%

Using the \TGTP\ project as a catalyst for this task I will try to
provide (working in conjunction with the authors of the tools):

\begin{itemize}
\item Converters from dynamic DGSs and GATPs tools (GCL language, Coq
  AM, etc.) to \itogatp\ format.

\item Converters from \itogatp\ format to DGSs and GATPs tools (GCL
  language, Coq AM, etc.).

\end{itemize}

The \itogatp format will be backwards compatible with \itog\ format.
DGSs should be able to read the \itogatp\ container ignoring the extra
info. The GATPs should also be able to read the \itog\ format, adding
information whenever needed.

The \TGTP\ and \GeoThms\ servers will use the \itogatp\ as its base
format, providing converters to and from the different GATPs.

\section{Conclusions and Further Work}
\label{sec:conclusions}

A case for extending the \itog\ {\XML} format to the description of
geometric conjectures and as an interchange format for dynamic
geometry software and geometry automated theorem proving tools was
presented.

A brief description of the \itog\ format and the tools using it and
also the tools that can benefit from the extended format was
given. The overall architecture and physical organisation of the
\itogatp\ format was described. Arguments justifying the usefulness of
this extended format were discussed.

The work presented in this paper is related to work in other domains of
automated reasoning where joint efforts of numerous researchers led to
standards and libraries which are very fruitful for easier exchange of
problems, proofs, and even program code, contributing to the advance of
the underlying field (see~\cite{Quaresma2011}).

This is a work-in-progress. Questions and future work to be addressed:

\begin{itemize}
\item The {\XML} format must be complemented with an extensive set of
  converters allowing the exchange of information between as many
  geometric tools as possible.

\item The databases queries, as in \TGTP, raise the question of
  selecting appropriate keywords. A fine grain index and/or an
  appropriate geometry ontology should be addressed.

\item The \itogatp\ format does not address proofs. Should we try to
  create such a format? The GATPs produce proofs in quite different
  formats, maybe the construction of such unifying format it is not
  possible and/or desirable in this area.
\end{itemize}

The \itogatp\ format will allow to further extend the database of
geometric constructions within \GeoThms\ and \TGTP\ and, hopefully lead
then to a major public resource for geometric constructions, linking a
significant number of geometry tools under this new format.

% ----------------------------------------------------

%\bibliographystyle{plain}
%\bibliography{pedro}

\begin{thebibliography}{10}

\bibitem{Chou2011}
Shang-Ching Chou, Xiao-Shan Gao, and Zheng Ye.
\newblock Java geometry expert.
\newblock \url{http://www.cs.wichita.edu/~ye/}, 2004.

\bibitem{Chou96}
Shang-Ching Chou, Xiao-Shan Gao, and Jing-Zhong Zhang.
\newblock Automated generation of readable proofs with geometric invariants,
  {I}. multiple and shortest proof generation.
\newblock {\em Journal of Automated Reasoning}, 17:325--347, 1996.

\bibitem{i2gImplementationTable}
The~Intergeo Consortium.
\newblock Intergeo implementation table.
\newblock \url{http://i2geo.net/xwiki/bin/view/I2GFormat/ImplementationsTable}.

\bibitem{Janicic06c}
Predrag Jani\v{c}i\'c.
\newblock {GCLC} — a tool for constructive euclidean geometry and more than
  that.
\newblock In Andr\'es Iglesias and Nobuki Takayama, editors, {\em Mathematical
  Software - ICMS 2006}, volume 4151 of {\em Lecture Notes in Computer
  Science}, pages 58--73. Springer Berlin / Heidelberg, 2006.

\bibitem{Janicic2010}
Predrag Jani\v{c}i\'c, Julien Narboux, and Pedro Quaresma.
\newblock The {A}rea {M}ethod: a recapitulation.
\newblock {\em Journal of Automated Reasoning}, 48(4):489--532, 2012.

\bibitem{Janicic06a}
Predrag Jani\v{c}i\'c and Pedro Quaresma.
\newblock System description: {GCLC}prover + {G}eo{T}hms.
\newblock In Ulrich Furbach and Natarajan Shankar, editors, {\em Automated
  Reasoning}, volume 4130 of {\em Lecture Notes in Computer Science}, pages
  145--150. Springer Berlin / Heidelberg, 2006.

\bibitem{Janicic06b}
Predrag Jani\v{c}i\'c and Pedro Quaresma.
\newblock Automatic verification of regular constructions in dynamic geometry
  systems.
\newblock In Francisco Botana and Tom\'as Recio, editors, {\em Automated
  Deduction in Geometry}, volume 4869 of {\em Lecture Notes in Computer
  Science}, pages 39--51. Springer Berlin / Heidelberg, 2007.

\bibitem{Kortenkamp2006}
U.~Kortenkamp, A.~M. Blessing, C.~Dohrmann, Y.~Kreis, P.~Libbrecht, and
  C.~Mercat.
\newblock Interoperable {I}nteractive {G}eometry for {E}urope – {F}irst
  technological and educational results and future challenges of the {I}ntergeo
  {P}roject.
\newblock In {\em CERME 6}, 2006.

\bibitem{Kortenkamp2009}
U.~Kortenkamp, C.~Dohrmann, Y.~Kreis, C.~Dording, P.~Libbrecht, and C.~Mercat.
\newblock Using the {I}ntergeo platform for teaching and research.
\newblock In {\em Proceedings of the 9th International Conference on Technology
  in Mathematics Teaching (ICTMT-9)}, 2009.

\bibitem{Matsuda04}
Noboru Matsuda and Kurt Vanlehn.
\newblock Gramy: A geometry theorem prover capable of construction.
\newblock {\em Journal of Automated Reasoning}, 32:3--33, 2004.

\bibitem{mathml2010}
W3C Math Working~Group members.
\newblock {\em Mathematical Markup Language (MathML) Version 3.0}.
\newblock W3C, October 2010.

\bibitem{Narboux04}
Julien Narboux.
\newblock A decision procedure for geometry in {C}oq.
\newblock {\em Lecture Notes in Computer Science}, 3223:225--240, 2004.

\bibitem{Narboux2007}
Julien Narboux.
\newblock A graphical user interface for formal proofs in geometry.
\newblock {\em Journal of Automated Reasoning}, 39:161--180, 2007.

\bibitem{Quaresma08}
P.~Quaresma, Toma\v{s}evi\'c~J. Jani\v{c}i\'c, P., M.~V.-Jani\v{c}i\'c, and
  D.~To\v{s}i\'c.
\newblock {\em Communicating Mathematics in The Digital Era}, chapter XML-Bases
  Format for Descriptions of Geometric Constructions and Proofs, pages
  183--197.
\newblock A. K. Peters, Ltd., 2008.

\bibitem{Quaresma2011}
Pedro Quaresma.
\newblock Thousands of geometric problems for geometric theorem provers (tgtp).
\newblock In Pascal Schreck, Julien Narboux, and Jürgen Richter-Gebert,
  editors, {\em Automated Deduction in Geometry}, volume 6877 of {\em Lecture
  Notes in Computer Science}, pages 169--181. Springer Berlin / Heidelberg,
  2011.

\bibitem{Quaresma2012a}
Pedro Quaresma.
\newblock The {\sc i2gatp} format.
\newblock Technical report, CISUC, 2012.
\newblock
  (\url{http://hilbert.mat.uc.pt/TGTP/Documents/Docs/cisucTRi2gatp.pdf}).

\bibitem{Quaresma2007}
Pedro Quaresma and Predrag Jani\v{c}i\'c.
\newblock {G}eo{T}hms -- a {W}eb {S}ystem for euclidean constructive geometry.
\newblock {\em Electronic Notes in Theoretical Computer Science}, 174(2):35 --
  48, 2007.

\bibitem{Santiago2010}
E.~Santiago, Maxim Hendriks, Yves Kreis, Ulrich Kortenkamp, and Daniel
  Marqu\`es.
\newblock {\sc i2g} {C}ommon {F}ile {F}ormat {F}inal {V}ersion.
\newblock Technical Report D3.10, The Intergeo Consortium, 2010.

\bibitem{Santos2012}
Vanda Santos and Pedro Quaresma.
\newblock Integrating {DGSs} and {GATPs} in an adaptative and collaborative
  blended-learning {W}eb-environment.
\newblock In {\em First Workshop on CTP Components for Educational Software
  (THedu'11)}, volume~79 of {\em EPTCS}, 2012.

\bibitem{Sutcliffe2008}
Geoff Sutcliffe.
\newblock The {SZS} ontologies for automated reasoning software.
\newblock In P~Rudnicki, G.~Sutcliffe, B.~Konev, R.~Schmidt, and S.~Schulz,
  editors, {\em Proceedings of the Combined KEAPPA - IWIL Workshops}, pages
  38--49, 2008.

\end{thebibliography}

\newcommand{\noopsort}[1]{} \newcommand{\singleletter}[1]{#1}

\end{document}